\documentclass[showpacs,preprintnumbers,amsmath,amssymb,twocolumn]{revtex4-1}
\usepackage{graphicx}% Include figure files
\usepackage{graphics}
\usepackage{dcolumn}% Align table columns on decimal point
\usepackage{bm}% bold math
\usepackage{subfigure}
\usepackage{float}
\usepackage{multirow}
\usepackage{booktabs}

\usepackage{amsmath}

\usepackage{latexsym,amssymb}
\usepackage[mathscr]{eucal}
\usepackage[switch*]{lineno}
\usepackage[bookmarks=true,
   colorlinks=true,
   linkcolor=blue,
   urlcolor=blue,
   citecolor=blue,
   bookmarks=true,
   hyperindex=true
]{hyperref}

\begin{document}

\title{Barrow entropy and stochastic gravitational wave background generated from cosmological QCD phase transition}
\author{Qi-Min Feng\textsuperscript{1,2}}
\author{Zhong-Wen Feng\textsuperscript{1,2}}
\altaffiliation{Email: zwfengphy@163.com}
\author{ Xia Zhou\textsuperscript{1}}
\author{Qing-Quan Jiang\textsuperscript{1,2}}
\vskip 0.5cm
\affiliation{1 School of Physics and Astronomy, China West Normal University, Nanchong, 637009, China \\
2 Institute of Theoretical Physics, China West Normal University, Nanchong, 637009, China}
%\date{\today}% It is always \today, today,
             %  but any date may be explicitly specified

\begin{abstract}
In this work we investigate the stochastic gravitational wave background generated during the f\/irst-order cosmological QCD phase transition of the early universe in the framework of the Barrow entropy. We f\/irst derived the Barrow corrections to the expression of stochastic gravitational wave background spectrum in presence of trace anomaly. Then, by taking account of Bubble wall collisions, sound waves and magnetohydrodynamic turbulence as the sources of stochastic gravitational wave, an analysis of the influence of Barrow entropy on the total energy density and the peak signal of stochastic gravitational wave signal is carried out. Finally, we discuss the possibility of detectors for the detection of these stochastic gravitational wave signals. Our results show that effect of Barrow entropy plays an important role in the temporal evolution of temperature of the universe as a function of the Hubble parameter, which leads to the signal of stochastic gravitational wave to shift towards the lower frequency regime,
making the signal of stochastic gravitational wave possible to be probed by relevant ongoing and upcoming gravitational waves experiments.
\end{abstract}
%\pacs{04.60.Bc, 04.80.Cc, 03.75.Dg}% PACS, the Physics and Astronomy
                             % Classification Scheme.
%Use showkeys class option if keyword
                              %display desired
\maketitle
% \linenumbers
\section{Introduction}
\label{intro}
The gravitational waves (GWs), as ripples in spacetimes, was first predicted by Albert Einstein in 1916, based on the theory of General Relativity (GR). Over the past one hundred years, physicists have done everything they can to detect the signal of gravitational waves, and hope it could improve people' understanding of the nature in the same way that electromagnetic waves have done. Fortunately, in 2015, the LIGO Scientif\/ic and Virgo Collaborations used sophisticated gravitational wave detectors to receive the signal of GWs from a binary black hole merger event for the first time \cite{cha1}. This GW events not only reaffirm the validity of GR, but also gives powerful evidence for the existence of black holes. Subsequently, more GWs signals from compact binary (such as black hole-neutron star \cite{cha2} and neutron star-neutron star \cite{cha3}) merger events have been detected. These results have made it possible to probe the mysteries of the universe from different perspectives, and have opened up the gravitational-wave astrophysics and multi-messenger astronomy. Now many more sensitive GWs detectors are being built and will be operational in the near future (see \cite{cha3+,cha4+,cha4,cha5,cha6,cha7,cha8,cha9} for example).

Despite the fact that GWs from compact binary merger events are the main focus today, it is also worth noting that GWs generated by other sources  also contain a wealth of unique physical information. In particular, many works  have been argued that the early universe have undergone several phase transitions during its evolution, if the phase transitions are strongly f\/irst order, they would have generated low-frequency GWs \cite{cha10,cha11,cha12}. Since the GWs propagate without interaction, at all, they would superimpose and populate the entire universe, and are therefore called stochastic gravitational wave background (SGWB). According to the standard model of particle physics, it is believed that there are at least two types of f\/irst-order cosmological phase transitions, viz. electroweak phase transition (at $T\sim100$ GeV) with accompanying electroweak symmetry breaking \cite{cha13,cha14,cha15}, and the quantum chromodynamics (QCD) phase transition (at $T\sim0.1-0.2$ GeV) that breaks the chiral symmetry \cite{cha16,cha17}.  In recent years, there has been an increasing interest in SGWB generated by QCD phase transitions, which is mainly due to two reasons: one is that the frequency of SGWB generated by QCD phase transitions is expected to be around $10^{-9}-10^{-6}$ Hz, which reaches the sensitivity of pulsar timing array experiments, such as the European Pulsar Timing Array (EPTA), International Pulsar Timing Array (IPTA), Square Kilometer Array (SKA), and North American Nanohertz Observatory for Gravitational Wave (NANOGrav). Therefore, it is believed that SGWB is likely to be detected in the near future. The other one is that the different materials and fields are able to influence the first-order phase transition in the QCD epoch and in turn change the features of the SGWB \cite{cha19,cha20,chc1,chc2,chc3,chc4,chc5}.  Since the signals of SGWB may contain traces of early universe, it is possible to use them to analyze the properties of the early universe and explore new physics within therein \cite{cha18,cha18+,cha18++,cha18+3}.

On the other hand, it has been argued that quantum gravity (QG) ef\/fects play a signif\/icant role in the generation of SGWB~\cite{chc6+,cha21,cha22,cha23,cha24,chc6}. For instance, the authors in~\cite{cha21,cha22,cha23} showed that the dif\/ferent kinds of generalized uncertainty principle (one of the most important QG model) can correct the thermal properties of the early universe, resulting in changes to the energy density and frequency of SGWB. Calcagni and Kuroyanagi \cite{cha24} analyzed the possibility of f\/ive cosmological QG scenarios generating SGWB, and discussed whether these signals could be detected by existing or future detectors. Subsequently, according to a very general class of ultraviolet complete theories of QG enjoying Weyl conformal invariance, Calcagni and Modesto \cite{chc6} constructed a new cosmological model and analyzed the properties of the GWs generated therein. Their results predict that these SGWB could be detected in the next 5-10 years.  Considering that faint QG effects are amplified as GWs propagate throughout the universe, the above works may provide a powerful probe into the production mechanisms of QG (see \cite{chc7} and references therein).

In 2020, based on the QG effects, Barrow \cite{cha28} proposed a possible fractal correction to the area law of the horizon entropy  (hereafter called the ``Barrow entropy", which can be considered as is a particular example of the new generalized entropy in~\cite{chy1,chy2}). The expression of  Barrow entropy is
 \begin{equation}
\label{eq1}
{S_B} = {\left( {\frac{A}{{4G}}} \right)^{1 + \frac{\delta }{2}}} = {\left( {{S_0}} \right)^{1 + \frac{\delta }{2}}},
\end{equation}
where $A$ is the standard horizon area, ${S_0}$  is the holographic entropy since its proportional to the boundary area of the system \cite{chb4}, the Barrow exponent $\delta  \in \left[ {0,1} \right]$ is a quantity indicating the effect of quantum gravitational deformation. For  $\delta=0$, the original expression of area-law entropy $S_0$ is recovered, which corresponds to the simplest horizon structure, whereas  $\delta=1$ indicates the most sophisticated surface structure. Notably,  Eq.~(\ref{eq1}) is very similar to that of  Tsallis entropy (${S_B}= \gamma A^\Delta$ with $\gamma$ is an unknown constant, and $\Delta\geq 0$ known as the Tsallis parameter) in non-generalized statistical thermodynamics \cite{chz1}, however, the two are fundamentally different in origin and motivation as well as physical principles. This interesting model has attracted a lot of attention since it was proposed, and the relevant applications have quickly expanded from the thermodynamics of black holes to the evolution of the universe \cite{cha29,cha30,cha31,cha32,cha33,cha34,cha36,cha37,cha38,chy3,chy4,chy5,chy6}. According to the above works, it is reason to believe that the Barrow entropy could  correct  the thermodynamic properties and evolution of the universe. Then some questions arise as whether the SGWB generated during the QCD phase transition epoch of the modif\/ied universe were af\/fected by the Barrow entropy? Are their signals being picked up by GW detectors?  To this end, we attempt to answer those questions in the present manuscript.  By utilizing the Barrow entropy~(\ref{eq1}), we derive the modified SGWB spectrum, the characteristic frequency, and the fractional energy density are analyzed. It turns out that the Barrow corrections to the peak of SGWB signals are lower than the original case, so that dif\/ferent detectors can detect SGWB signals with different  $\delta$.

The paper is organized as follows. In the next section, according to the Barrow entropy, we derive the general expression for the SGWB spectrum measured today. In section~\ref{sec3}, by comparing the equation of state phase of the ideal gas with the QCD equation of state provided by recent lattice results, we obtain the expression for the ef\/fective equation of state used in this study. In Section~\ref{sec4}, by discussing the role of  Barrow entropy on the amplitude and frequency of  SGWB,  we predict the response of the modif\/ied SGWB signs in ongoing and proposed detectors for future observations. Finally, the conclusion is given in  section~\ref{sec5}.

\section{Modified SGWB spectrum from Barrow corrections}
\label{sec2}
\begin{figure}[htbp]
\centering
\includegraphics[width=0.5\textwidth]{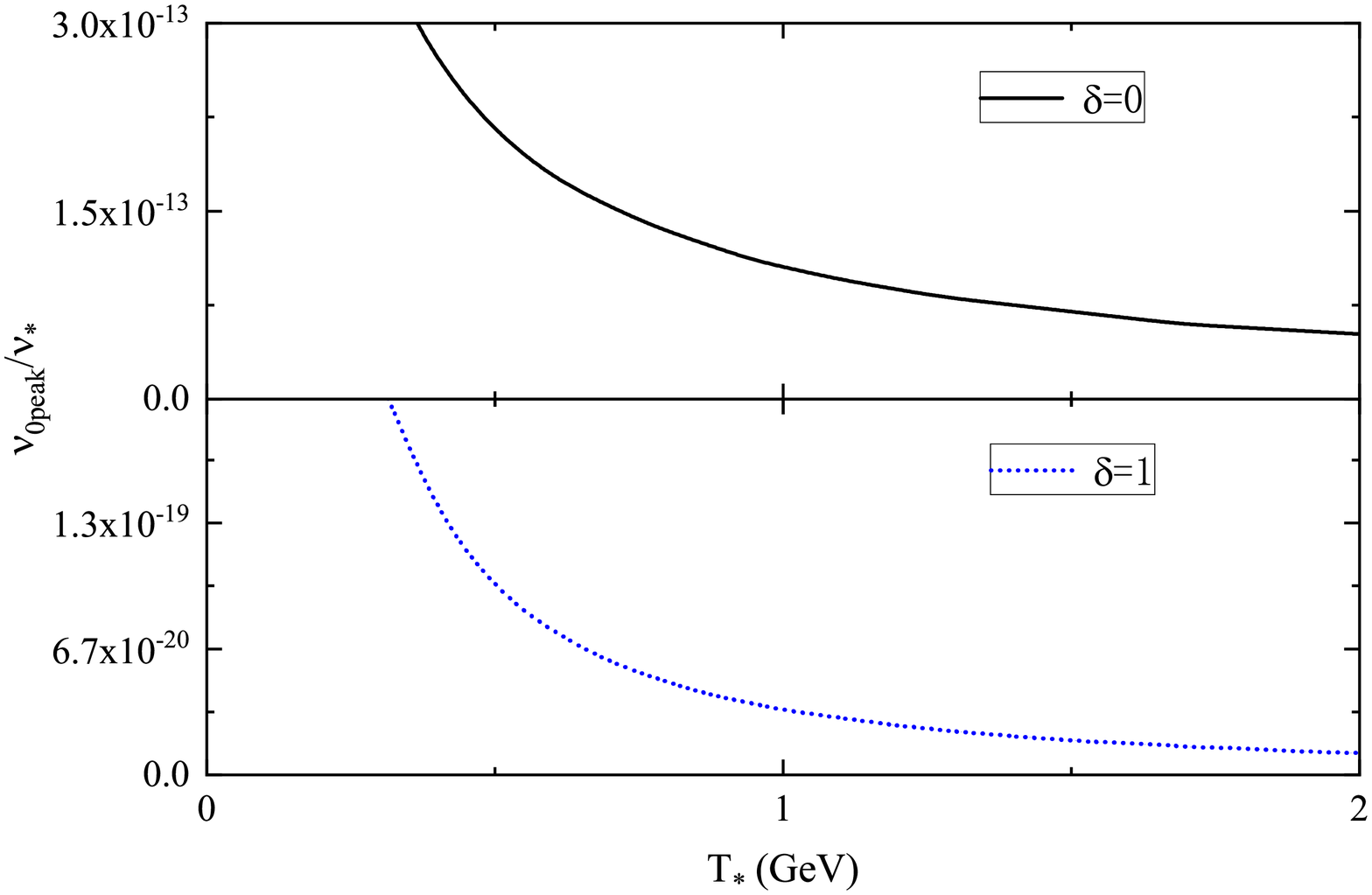}
\caption{\label{fig1} ${\nu _{0{\text{peak}}}}/{\nu _*}$  for dif\/ferent values of Barrow exponent $\delta$  with  ${T_0} = 2.725 K = 2.348 \times {10^{ - 13}}{\text{GeV}}$, ${g_s}\left( {{T_0}} \right) = 3.4$  and  ${g_s}\left( {{T_*}} \right) = 35$.}
\end{figure}

A consensus was reached that SGWB is generated during the QCD and electroweak phase transition periods, and propagated to the current epoch. Therefore, for investigating the SGWB today, we consider the GWs from the epoch of phase transition to the present time. By assuming that the universe is expanding adiabatically due to the phase transition, the entropy density can be estimated to be $s \sim {a^3} {S_0}\sim {a^3}\xi {g_s}{T^3}$, where $a$,   ${g_s}$ and $\xi$ represent the scale factor, the ef\/fective number of degrees of freedom involved in entropy density and an undetermined parameter, respectively \cite{cha39}. Now, applying the Barrow entropy~(\ref{eq1}),  the relevant entropy density becomes
\begin{align}
\label{eq2}
{s_B}={a^3} {\left( {\frac{{2{\pi ^2}}}{{45}}{g_\pi }{T^3}} \right)^{1 + \frac{\delta }{2}}},
\end{align}
where we take $\xi  = {{2{\pi ^2}} \mathord{\left/ {\vphantom {{2{\pi ^2}} {45}}} \right. \kern-\nulldelimiterspace} {45}}$  in accordance with the results in~\cite{cha22}. Since the entropy density in the adiabatic expanding universe remains constant even beyond equilibrium (i.e.,  ${{\dot s} \mathord{\left/ {\vphantom {{\dot s} s}} \right. \kern-\nulldelimiterspace} s} = 0$), the time variation of universe temperature can be expressed as
 \begin{align}
\label{eq3}
\frac{{{\text{d}}T}}{{{\text{d}}t}} =  - HT\Xi {\left( {T,{g_s},\delta } \right)^{ - 1}},
\end{align}
where  $H$  is the Hubble parameter, and $\Xi \left( {T,{g_s},\delta } \right) = 1 + \frac{T}{{3{g_s}}}\frac{{{\text{d}}{g_s}}}{{{\text{d}}T}} + \frac{\delta }{2}\left( {1 + \frac{T}{{3{g_s}}}\frac{{{\text{d}}{g_s}}}{{{\text{d}}T}}} \right)$. It is clear that the original case $\frac{{{\text{d}}T}}{{{\text{d}}t}} =  - HT{\left( {1 + \frac{T}{{3{g_s}}}\frac{{{\text{d}}{g_s}}}{{{\text{d}}T}}} \right)^{ - 1}}$ is recovered when  $\delta=0$. Next, due to the Hubble parameter, one can rewrite Eq.~(\ref{eq3}) in terms of scale factor
 \begin{align}
\label{eq4}
\frac{{{a_*}}}{{{a_0}}} = \exp \left[ {\int_{{T_*}}^{{T_0}} {\frac{1}{T}\Xi \left( {T,{g_s},\delta } \right)} {\text{d}}T} \right].
\end{align}
Here and also in what follows, the subscripts ``$*$" and ``$0$" denote the quantities at the epochs of phase transition and today, respectively. Furthermore, based on the relation between the scale factor and red-shifted  ${{{\nu _{0{\text{peak}}}}} \mathord{\left/ {\vphantom {{{\nu _{0{\text{peak}}}}} {{\nu _*}}}} \right. \kern-\nulldelimiterspace} {{\nu _*}}} = {{{a_*}} \mathord{\left/ {\vphantom {{{a_*}} {{a_0}}}} \right. \kern-\nulldelimiterspace} {{a_0}}}$, the peak frequency of SGWB red-shifted to current epoch is given by
\begin{align}
\label{eq5}
{ {\frac{{{\nu _{0{\text{peak}}}}}}{{{\nu _*}}}} } = { {\frac{{{a_*}}}{{{a_0}}}}} = \frac{{{T_0}}}{{{T_*}}}{\left[ {\frac{{{g_s}\left( {{T_0}} \right)}}{{{g_s}\left( {{T_*}} \right)}}} \right]^{\frac{{2 + \delta }}{6}}}{\left( {\frac{{{T_0}}}{{{T_*}}}} \right)^{\frac{\delta }{2}}}.
\end{align}
In order to gain insight into the ef\/fect of Barrow entropy on ${{{\nu _{0{\text{peak}}}}} \mathord{\left/ {\vphantom {{{\nu _{0{\text{peak}}}}} {{\nu _*}}}} \right. \kern-\nulldelimiterspace} {{\nu _*}}}$, we depicted Fig.~\ref{fig1}. It is found that dif\/ferent values of Barrow exponent $\delta$ give a similar behavior for the ratio  ${{{\nu _{0{\text{peak}}}}} \mathord{\left/ {\vphantom {{{\nu _{0{\text{peak}}}}} {{\nu _*}}}} \right. \kern-\nulldelimiterspace} {{\nu _*}}}$, that is, the ${{{\nu _{0{\text{peak}}}}} \mathord{\left/ {\vphantom {{{\nu _{0{\text{peak}}}}} {{\nu _*}}}} \right. \kern-\nulldelimiterspace} {{\nu _*}}}$ increases as the temperature  $T _*$ decreases. Meanwhile, when the $T_*$ is fixed, the ratio ${{{\nu _{0{\text{peak}}}}} \mathord{\left/ {\vphantom {{{\nu _{0{\text{peak}}}}} {{\nu _*}}}} \right. \kern-\nulldelimiterspace} {{\nu _*}}}$ decreases significantly with increasing $\delta$.

Next, considering the SGWB is ultimately decoupled to the rest of the universe, one can derive the energy density of the SGWB at today epoch by employing the Boltzmann equation  ${{{\text{d}}\left( {{\rho _{{\text{gw}}}}{a^4}} \right)} \mathord{\left/ {\vphantom {{{\text{d}}\left( {{\rho _{{\text{gw}}}}{a^4}} \right)} {{\text{d}}t}}} \right.
 \kern-\nulldelimiterspace} {{\text{d}}t}} = 0$, which yields
\begin{align}
\label{eq6}
{\rho _{{\text{gw}}}}\left( {{T_0}} \right) &  = {\rho _{{\text{gw}}}}\left( {{T_*}} \right){\left( {\frac{{{a_*}}}{{{a_0}}}} \right)^4}
\nonumber \\
&   = {\rho _{{\text{gw}}}}\left( {{T_*}} \right) \exp \left[ {\int_{{T_*}}^{{T_0}} {\frac{4}{T}\Xi \left( {T,{g_s},\delta } \right){\text{d}}T} } \right].
\end{align}
Then, by using the def\/initions of the density parameter of SGWB in the today's epoch ${\Omega _{{\text{gw}}}} = {{{\rho _{{\text{gw}}}}\left( {{T_0}} \right)} \mathord{\left/ {\vphantom {{{\rho _{{\text{gw}}}}\left( {{T_0}} \right)} {{\rho _{{\text{cr}}}}\left( {{T_0}} \right)}}} \right. \kern-\nulldelimiterspace} {{\rho _{{\text{cr}}}}\left( {{T_0}} \right)}}$  as well as its counterpart in phase transition era  ${\Omega _{{\text{gw}}*}} = {{{\rho _{{\text{gw}}}}\left( {{T_*}} \right)} \mathord{\left/  {\vphantom {{{\rho _{{\text{gw}}}}\left( {{T_*}} \right)} {{\rho _{{\text{cr}}}}\left( {{T_*}} \right)}}} \right. \kern-\nulldelimiterspace} {{\rho _{{\text{cr}}}}\left( {{T_*}} \right)}}$, the expression of density parameter of SGWB in today's epoch is given by
 \begin{align}
\label{eq7}
{\Omega _{{\text{gw}}}} = {\Omega _{{\text{gw}}*}}{\left( {\frac{{{H_*}}}{{{H_0}}}} \right)^2}\exp \left[ {\int_{{T_*}}^{{T_0}} {\frac{4}{T}\Xi \left( {T,{g_s},\delta } \right){\text{d}}T} } \right],
\end{align}
where
\begin{align}
\label{eq8}
{\left( {\frac{{{H_*}}}{{{H_0}}}} \right)^2} = \frac{{{\rho _{{\text{cr}}}}\left( {{T_*}} \right)}}{{{\rho _{{\text{cr}}}}\left( {{T_0}} \right)}}.
\end{align}

In order to obtain the ratio of the Hubble parameters during the phase transition to its current value, it is necessary to apply the continuity equation ${\dot \rho _t} =  - 3H{\rho _t}\left( {1 + {{{P_t}} \mathord{\left/  {\vphantom {{{P_t}} {{\rho _t}}}} \right. \kern-\nulldelimiterspace} {{\rho _t}}}} \right)$ with the total pressure density of the universe $P_t$  and total energy density of the universe $\rho_t$. Inserting Eq.~(\ref{eq3}) into the continuity equation, one has
\begin{align}
\label{eq9}
\frac{{{\text{d}}{\rho _t}}}{{{\rho _t}}} = \frac{3}{T}\left( {1 + {\omega _{{\text{ef\/f}}}}} \right)\Xi \left( {T,{g_s},\delta } \right){\text{d}}T.
\end{align}
where ${\omega _{{\text{ef\/f}}}} = {{{P_t}} \mathord{\left/  {\vphantom {{{P_t}} {{\rho _t}}}} \right. \kern-\nulldelimiterspace} {{\rho _t}}}$  is the effective equation of state parameter. By integrating Eq.~(\ref{eq9}) from the early time in the radiation dominated  ${T_r} = {10^4}$ GeV to the time of phase transition  ${T_*}$, the critical energy density of radiation in the phase transition period is given by
\begin{align}
\label{eq10}
{\rho _{cr}}\left( {{T_*}} \right) = {\rho _r}\left( {{T_r}} \right) \exp \left[ {\int_{{T_r}}^{{T_*}} {\frac{3}{T}\left( {1 + {\omega _{{\text{ef\/f}}}}} \right)\Xi \left( {T,{g_s},\delta } \right){\text{d}}T} } \right].
\end{align}
Then, substituting Eq.~(\ref{eq10}) into Eq.~(\ref{eq8}), and using the relation  $H_*^2 = {\rho _*}$ \cite{chd1}, one gets
\begin{align}
\label{eq11}
 {\left( {\frac{{{H_*}}}{{{H_0}}}} \right)^2} & = {\Omega _{r0}}\frac{{{\rho _r}\left( {{T_r}} \right)}}{{{\rho _r}\left( {{T_0}} \right)}}
\nonumber \\
& \times \exp \left[ {\int_{{T_r}}^{{T_*}} {\frac{3}{T}\left( {1 + {\omega _{{\text{ef\/f}}}}} \right)\Xi \left( {T,{g_s},\delta } \right){\text{d}}T} } \right],
\end{align}
where ${\Omega _{r0}} = {{{\rho _r}\left( {{T_0}} \right)} \mathord{\left/ {\vphantom {{{\rho _r}\left( {{T_0}} \right)} {{\rho _{cr}}\left( {{T_0}} \right)}}} \right. \kern-\nulldelimiterspace} {{\rho _{cr}}\left( {{T_0}} \right)}} \simeq 8.5 \times {10^{ - 5}}$  represents the current value of fractional energy density of radiation. Moreover, according to the viewpoint in~\cite{cha23}, it is proved that  ${{{\rho _r}\left( {{T_r}} \right)} \mathord{\left/
 {\vphantom {{{\rho _r}\left( {{T_r}} \right)} {{\rho _r}\left( {{T_0}} \right)}}} \right. \kern-\nulldelimiterspace} {{\rho _r}\left( {{T_0}} \right)}} \simeq {\left( {{{{a_0}} \mathord{\left/ {\vphantom {{{a_0}} {{a_r}}}} \right. \kern-\nulldelimiterspace} {{a_r}}}} \right)^4}$, hence, Eq.~(\ref{eq11}) can be rewritten as
\begin{align}
\label{eq12}
{\left( {\frac{{{H_*}}}{{{H_0}}}} \right)^2} & = {\Omega _{r0}}\exp \left[ {\int_{{T_0}}^{{T_r}} {\frac{4}{T}\Xi \left( {T,{g_s},\delta } \right){\text{d}}T} } \right]
\nonumber \\
& \times \exp \left[ {\int_{{T_r}}^{{T_*}} {\frac{3}{T}\left( {1 + {\omega _{{\text{ef\/f}}}}} \right)\Xi \left( {T,{g_s},\delta } \right)){\text{d}}T} } \right].
\end{align}
 As a result, we find the GW spectrum measured today, from Eq.~(\ref{eq7}), as
\begin{align}
\label{eq13}
{\Omega _{{\text{gw}}}} & = {\Omega _{r0}}{\Omega _{{\text{gw}}*}}\exp \left[ {\int_{{T_*}}^{{T_r}} {\frac{4}{T}\Xi \left( {T,{g_s},\delta } \right){\text{d}}T} } \right]
\nonumber \\
& \times\exp \left[ {\int_{{T_r}}^{{T_*}} {\frac{3}{T}\left( {1 + {\omega _{{\text{ef\/f}}}}} \right)\Xi \left( {T,{g_s},\delta } \right){\text{d}}T} } \right].
\end{align}
From Eq.~(\ref{eq12}) and Eq.~(\ref{eq13}), it is found that the ratio ${{{H_*}} \mathord{\left/ {\vphantom {{{H_*}} {{H_0}}}} \right. \kern-\nulldelimiterspace} {{H_0}}}$  and  ${\Omega _{{\text{gw}}}}$  are related to the ef\/fective equation of state parameter ${\omega_\text{ef\/f}}$  and the Barrow exponent  $\delta$. Therefore, in the following content, we will discuss the inf\/luence of these two parameters on the generation of SGWB arising from the QCD phase transition.

\section{The equation of state of  SGWB}
\label{sec3}
In this section, it is necessary to analyze the inf\/luence of the equation of state on SGWB. For the ultra-relativistic ideal gas with non-interacting particle, the effect equation of state is equal to  ${\omega _{{\text{ef\/f}}}} = {1 \mathord{\left/ {\vphantom {1 3}} \right. \kern-\nulldelimiterspace} 3}$, which leads to Eq.~(\ref{eq12}) and Eq.~(\ref{eq13}) turning into the following forms
 \begin{align}
\label{eq14}
{\left( {\frac{{{H_*}}}{{{H_0}}}} \right)^2} = {\Omega _{r0}}{\left( {\frac{{{T_*}}}{{{T_0}}}} \right)^{2\left( {2 + \delta } \right)}}{\left[ {\frac{{{g_s}\left( {{T_*}} \right)}}{{{g_s}\left( {{T_0}} \right)}}} \right]^{\frac{{2\left( {2 + \delta } \right)}}{3}}},
\end{align}
and
 \begin{align}
\label{eq15}
{\Omega _{{\text{gw}}}} = {\Omega _{r0}}{\Omega _{{\text{gw}}*}}.
\end{align}
Obviously, in the ultra-relativistic gas with non-interacting particle case, Eq.~(\ref{eq12}) and Eq.~(\ref{eq13}) become very concise. In particular, the SGWB spectrum measured today~(\ref{eq15}) reduces to the original scenario, which means the Barrow exponent $\delta$  is no longer affecting SGWB  spectrum measured today. However, QCD interactions can lead to deviation from ${\omega _{{\text{ef\/f}}}} = {1 \mathord{\left/ {\vphantom {1 3}} \right. \kern-\nulldelimiterspace} 3}$ ~\cite{cha39}, resulting in SGWB that are significantly dif\/ferent from those in the ideal gas case. Therefore, we explore the equation of state  around QCD epoch.

\begin{figure}[htbp]
\centering
\includegraphics[width=0.5\textwidth]{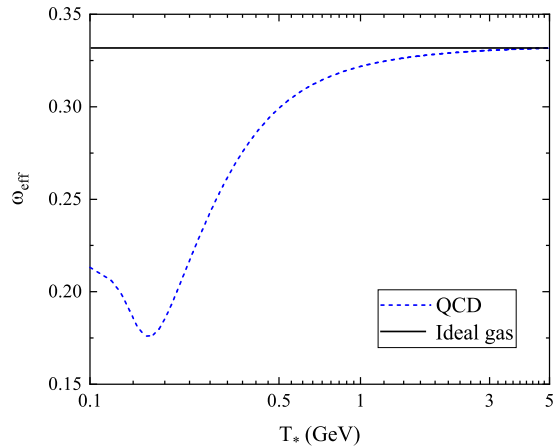}
\caption{\label{fig2} The effective equation of state ${\omega _{{\text{ef\/f}}}}$   as a  function of  transition temperature  ${T_*}$.}
\end{figure}

In~\cite{cha40}, the authors pointed out that the ef\/fect of QCD interaction can be introduced by using the results of modern lattice calculation, which employs the ${N_f} = 2 + 1$  f\/lavours and covers the temperature  $0.1- 0.4$ GeV. The QCD equation of state resulting from the pressure parameterization due to the strong interactions between $u$, $d$, $s$ quarks and gluons, as
 \begin{align}
\label{eq16}
& F\left( T \right)  = \frac{P}{{{T^4}}}
 \nonumber \\
 &=  \frac{1}{2}\left\{ {1 + \tanh \left[ {{c_\tau }\left( {\tau  - {\tau _0}} \right)} \right]} \right\} \frac{{{p_i} + \frac{{{a_n}}}{\tau } + \frac{{{b_n}}}{{{\tau ^2}}} + \frac{{{c_n}}}{{{\tau ^4}}}}}{{1 + \frac{{{a_d}}}{\tau } + \frac{{{b_d}}}{{{\tau ^2}}} + \frac{{{c_d}}}{{{\tau ^4}}}}},
\end{align}
where $\tau  = {T \mathord{\left/ {\vphantom {T {{T_c}}}} \right. \kern-\nulldelimiterspace} {{T_c}}}$  with the phase transition temperature  ${T_c} = 0.145$ GeV, ${p_i} = {{19{\pi ^2}} \mathord{\left/ {\vphantom {{19{\pi ^2}} {36}}} \right. \kern-\nulldelimiterspace} {36}}$  represents the ideal gas value of  ${P \mathord{\left/ {\vphantom {P {{T^4}}}} \right. \kern-\nulldelimiterspace} {{T^4}}}$ for QCD with three massless quarks. The other coef\/f\/icients, at temperatures above 100 MeV, can be expressed as  ${c_\tau } = 3.6706$,  ${\tau _0} = 0.9761$,  ${a_n} =  - 8.7704$,  ${b_n} = 3.9200$,  ${c_n} = 0.3419$,  ${a_d} =  - 1.2600$,  ${b_d} = 0.8425$,  ${c_d} =  - 0.0475$. Moreover, according to Eq.~(\ref{eq16}), the trace anomaly can be expressed in terms of energy density and pressure as \cite{cha41}
 \begin{align}
\label{eq17}
 \frac{{\rho  - 3P}}{{{T^4}}}  = T\frac{{{\text{d}}F\left( T \right)}}{{{\text{d}}T}} = T\frac{{\text{d}}}{{{\text{d}}T}}\left( {\frac{P}{{{T^4}}}} \right),
\end{align}
and the ef\/fective equation of state with trace anomaly effect is given by
\begin{align}
\label{eq18}
{\omega _{{\text{eff}}}} = {\left[ {\frac{T}{{F\left( T \right)}}\frac{{{\text{d}}F\left( T \right)}}{{{\text{d}}T}} + 3} \right]^{ - 1}}.
\end{align}
In order to probe the role of trace anomaly ef\/fect on the ef\/fective equation of state, we display Fig.~\ref{fig2}.

From Fig.~\ref{fig2}, one can see  a very large gap between the blue dashed curve for the ef\/fective equation of state with trace anomaly ef\/fect and black solid curve for the ideal gas ${\omega _{{\text{ef\/f}}}}$  around the critical temperature region ($0.1-0.2$ GeV), which indicates the ef\/fect of trace anomaly around the QCD phase transition epoch cannot be neglected. However, at the high temperatures (${T_*} \geq 5$ GeV), the two curves converge to each other and hence are asymptotically equal.

\begin{figure}[htbp]
\centering
\subfigure[]{
\includegraphics[scale=0.31]{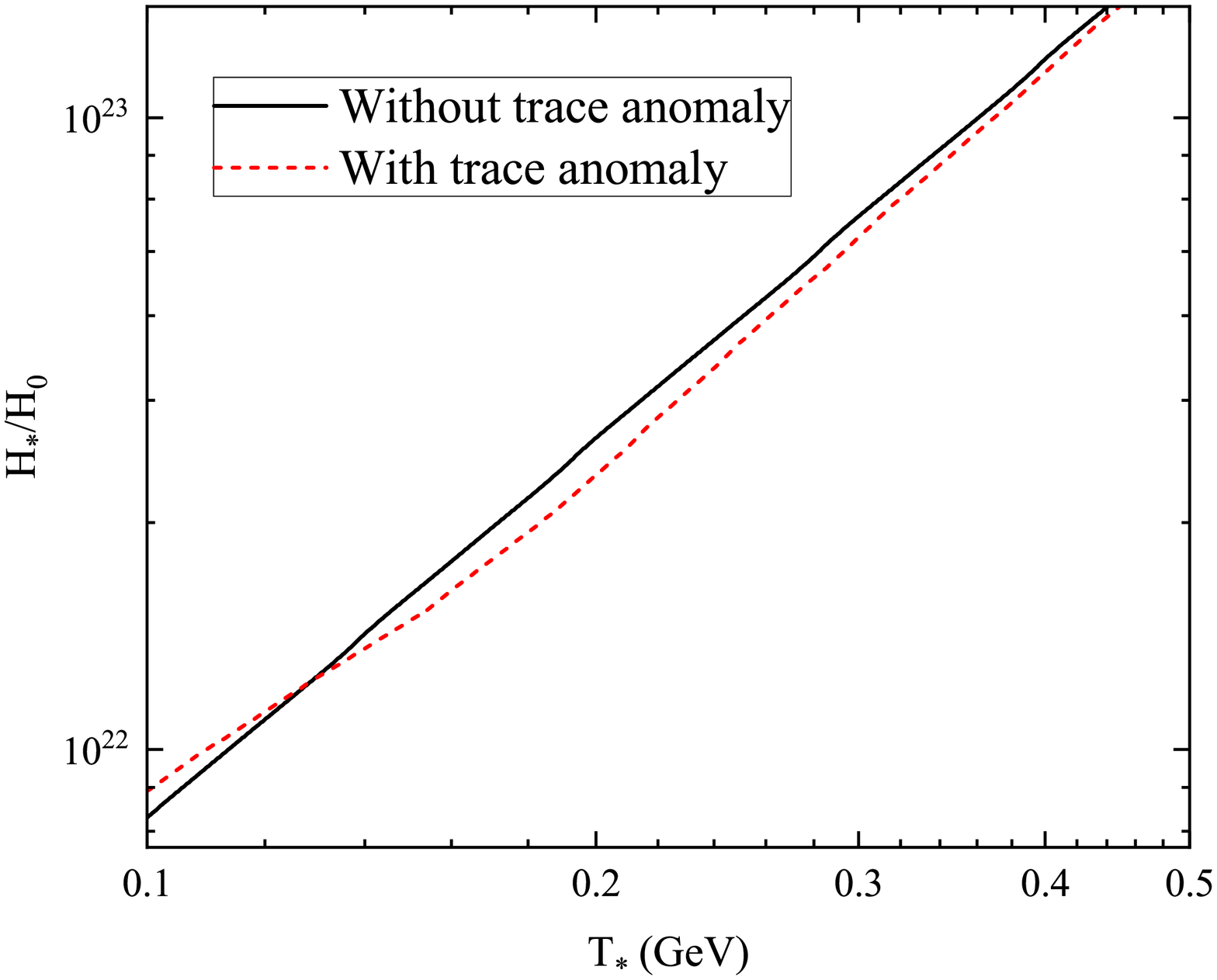}
\label{fig3-a}
}
\quad
\subfigure[]{
\includegraphics[scale=0.31]{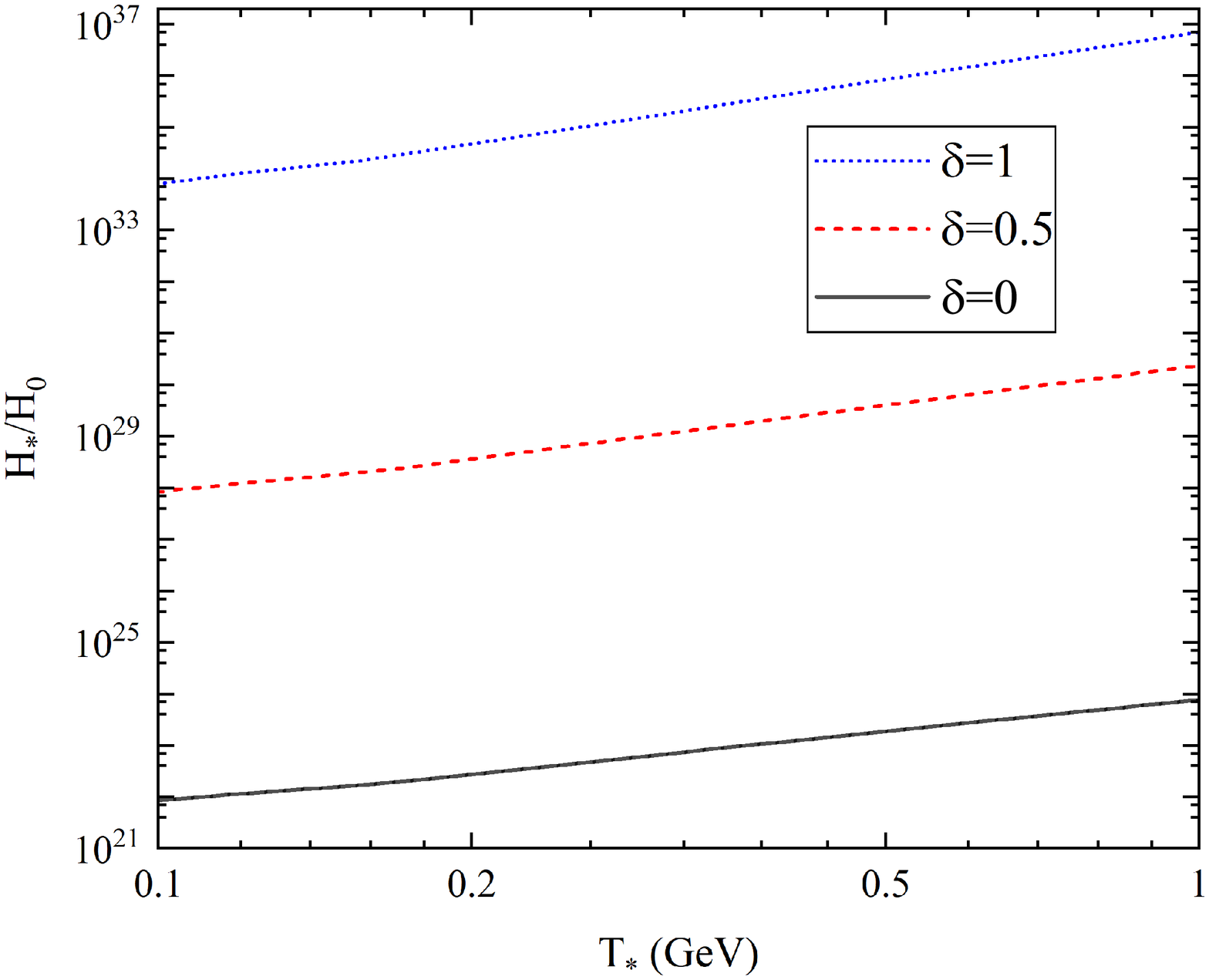}
\label{fig3-b}
}
\caption{(a) The ${H_*}/{H_0}$  as a function of transition temperature  ${T_*}$ without Barrow exponent.
(b) The  ${H_*}/{H_0}$  as a function of transition temperature  ${T_*}$ for dif\/ferent Barrow exponent. We set  ${T_r} = {10^4}{\text{GeV}}$.}
\label{fig3}
\end{figure}

By putting Eq.~(\ref{eq18}) into Eq.~(\ref{eq13}), the ratio between Hubble parameter at the epoch of transition to its counterpart at current epoch ${\left( {{{{H_*}} \mathord{\left/ {\vphantom {{{H_*}} {{H_0}}}} \right. \kern-\nulldelimiterspace} {{H_0}}}} \right)^2}$  takes the form
\begin{align}
\label{eq19}
{\left( {\frac{{{H_*}}}{{{H_0}}}} \right)^2} & = {\Omega _{r0}}{\left( {\frac{{{T_r}}}{{{T_0}}}} \right)^{2\left( {2 + \delta } \right)}}{\left[ {\frac{{{g_s}\left( {{T_r}} \right)}}{{{g_s}\left( {{T_0}} \right)}}} \right]^{\frac{{2\left( {2 + \delta } \right)}}{3}}}{\left[ {\frac{{{g_s}{{\left( {{T_*}} \right)}^{1 + \omega \left( {{T_*}} \right)}}}}{{{g_s}{{\left( {{T_r}} \right)}^{1 + \omega \left( {{T_*}} \right)}}}}} \right]^{\frac{{2 + \delta }}{2}}}
  \nonumber \\
&   \times \exp \left[ {\int_{{T_r}}^{{T_*}} {\frac{3}{T}\left( {1 + {\omega _{{\text{ef\/f}}}}} \right)\left( {1 + \frac{\delta }{2}} \right){\text{d}}T} } \right].
\end{align}
According to the above equation, we plot the  ${\left( {{{{H_*}} \mathord{\left/ {\vphantom {{{H_*}} {{H_0}}}} \right. \kern-\nulldelimiterspace} {{H_0}}}} \right)^2}$  versus transition temperature  ${T_*}$ in Fig.~\ref{fig3}.

In Fig.~\ref{fig3-a}, we show the  ${{{H_*}} \mathord{\left/ {\vphantom {{{H_*}} {{H_0}}}} \right. \kern-\nulldelimiterspace} {{H_0}}}$  as a function of transition temperature ${T_*}$  with  $\delta=0$. The black solid curve represents ${{{H_*}} \mathord{\left/ {\vphantom {{{H_*}} {{H_0}}}} \right. \kern-\nulldelimiterspace} {{H_0}}}$  without trace anomaly, whereas the trace anomaly case is illustrated by the red dashed curve. It can be seen that the slope of the black solid curve remains constant and its rises uniformly with the increase in transition temperature ${T_*}$. However, when considering the ef\/fect of the trace anomaly,  the slope of red dashed curve is significantly lower than that of black solid curve in the region ${T_*} \leq 0.13$ GeV.  For  ${T_*} > 0.13$ GeV, the slope of the red dashed curve gets raised, which it catches up with black solid curve at  ${T_*} \approx 5$ GeV. It is again shown that in order to study the SGWB generated from cosmological QCD phase transition epoch, the effect of trace anomaly must be taken into account. Therefore,  we will consider the ef\/fect of trace anomaly in the followings. From Fig.~\ref{fig3-b}, by taking account the effect of trace anomaly, one can see the relationship between  ${{{H_*}} \mathord{\left/ {\vphantom {{{H_*}} {{H_0}}}} \right. \kern-\nulldelimiterspace} {{H_0}}}$  and transition temperature ${T_*}$  for different Barrow exponent. The value of Barrow exponent  decreases from top (red dashed curve) to bottom (blue solid curve). It is clear that the ratio ${{{H_*}} \mathord{\left/ {\vphantom {{{H_*}} {{H_0}}}} \right. \kern-\nulldelimiterspace} {{H_0}}}$  with different Barrow exponent $\delta$ have a similar behavior with changing transition temperature ${T_*}$.  If fixing the transition temperature, the ratio increases with increasing Barrow exponent.

\begin{figure}[htbp]
\centering
\includegraphics[width=0.5\textwidth]{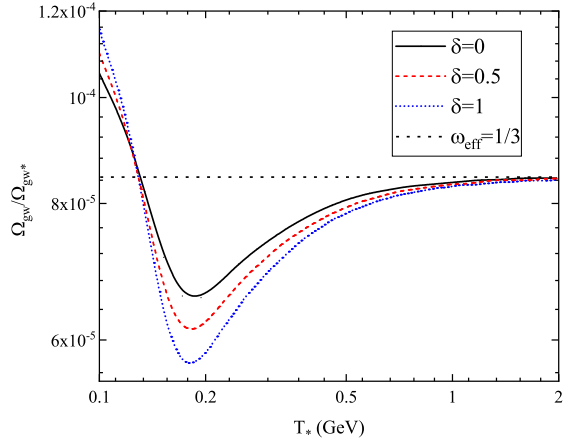}
\caption{\label{fig4} The plot of the relationship between  ${\Omega _{{\text{gw}}}}/{\Omega _{{\text{gw}}*}}$  with respect to the function of transition temperature  ${T_*}$ for different $\delta $.}
\end{figure}

Furthermore, applying the ef\/fect of trace anomaly and Eq.~(\ref{eq13}), the ratio ${{{\Omega _{{\text{gw}}}}} \mathord{\left/ {\vphantom {{{\Omega _{{\text{gw}}}}} {{\Omega _{{\text{gw}}*}}}}} \right. \kern-\nulldelimiterspace} {{\Omega _{{\text{gw}}*}}}}$  can be rewritten as
\begin{align}
\label{eq20}
  \frac{{{\Omega _{{\text{gw}}}}}}{{{\Omega _{{\text{gw}}*}}}} & = {\Omega _{r0}}{\left( {\frac{{{T_r}}}{{{T_*}}}} \right)^{2\left( {2 + \delta } \right)}}{\left[ {\frac{{{g_s}\left( {{T_r}} \right)}}{{{g_s}\left( {{T_*}} \right)}}} \right]^{\frac{2}{3}\left( {2 + \delta } \right)}}{\left[ {\frac{{{g_s}{{\left( {{T_*}} \right)}^{1 + \omega \left( {{T_*}} \right)}}}}{{{g_s}{{\left( {{T_r}} \right)}^{1 + \omega \left( {{T_r}} \right)}}}}} \right]^{\frac{{2 + \delta }}{2}}}
\nonumber \\
& \times \exp \left[ {\int_{{T_r}}^{{T_*}} {\frac{3}{T}\left( {1 + {\omega _{{\text{ef\/f}}}}} \right)\left( {1 + \frac{\delta }{2}} \right){\text{d}}T} } \right].
\end{align}
By setting ${T_r} = {10^4}$ GeV  and  ${g_s}\left( {{T_r}} \right) = 106$, the relationship between ${{{\Omega _{{\text{gw}}}}} \mathord{\left/ {\vphantom {{{\Omega _{{\text{gw}}}}} {{\Omega _{{\text{gw}}*}}}}} \right. \kern-\nulldelimiterspace} {{\Omega _{{\text{gw}}*}}}}$  as a function of transition temperature ${T_*}$  for dif\/ferent  $\delta$ is displayed in Fig.~\ref{fig4}.

In Fig.~\ref{fig4}, the black horizontal line represents the equation of state for ultra-relativistic gas with non-interacting particles, i.e., ${\omega _{{\text{ef\/f}}}} = {1 \mathord{\left/ {\vphantom {1 3}} \right. \kern-\nulldelimiterspace} 3}$. The black solid curve, red dashed curve, and blue dotted curve illustrate the ${{{\Omega _{{\text{gw}}}}} \mathord{\left/ {\vphantom {{{\Omega _{{\text{gw}}}}} {{\Omega _{{\text{gw}}*}}}}} \right. \kern-\nulldelimiterspace} {{\Omega _{{\text{gw}}*}}}}$  with  $\delta=0$, $0.5$ and $1$, respectively. Before ${T_*} \leq 0.13{\text{GeV}}$, the  ${{{\Omega _{{\text{gw}}}}} \mathord{\left/ {\vphantom {{{\Omega _{{\text{gw}}}}} {{\Omega _{{\text{gw}}*}}}}} \right. \kern-\nulldelimiterspace} {{\Omega _{{\text{gw}}*}}}}$ inf\/luenced by Barrow exponent is higher than the original case, whereas it is below the original case when  ${T_*} > 0.13{\text{GeV}}$. In the high temperature region, the QCD equation of state becomes the case of ultra-relativistic gas with non-interacting particles, which leads to all the curves intersect the black dotted line at $2$ GeV .

\section{Modified QCD sources of SGWB}
\label{sec4}
During the f\/irst-order phase transitions, the true vacuum bubbles nucleate and expand in the false vacuum. Collisions of these bubbles create a quadrupole moment in the cosmic f\/luid when they occur and lead to the generation of SGWB.  Generally, the dominant components of the SGWB signal are: (i) the collisions of bubble wall collisions \cite{che1}, (ii) the sound waves \cite{che2}, and (iii) the magnetohydrodynamic turbulence which produced after bubble collision in the plasma \cite{che3,che4}. Therefore, the contributions from the three sources to  ${\Omega _{{\text{gw}}*}} = {\Omega _{\text{gw}}}\left( T_* \right)$, respectively, can be given by:
\\(a) Bubble wall collisions (BC)
\begin{align}
\label{eq21}
\Omega _{{\text{gw}}*}^{{\text{BC}}}(\nu)  = {\left( {\frac{{{H_*}}}{\beta }} \right)^2}{\chi _{{\text{BC}}}}\left( {\frac{{0.11{\mu ^3}}}{{0.42 + {\mu ^2}}}} \right){S_{{\text{BC}}}},
{\chi _{{\text{BC}}}} = {\left( {\frac{{{\kappa _{{\text{BC}}}}}}{{1 + \epsilon}}} \right)^2}.
\end{align}
\\(b) Sound waves (SW)
\begin{align}
\label{eq22}
\Omega _{{\text{gw}}*}^{{\text{SW}}}(\nu )  = {\left( {\frac{{{H_*}}}{\beta }} \right)^2}{\chi _{{\text{SW}}}}\mu {S_{{\text{SW}}}},
{\chi _{{\text{SW}}}} = {\left( {\frac{{{\kappa _{{\text{SW}}}}}}{{1 + \epsilon}}} \right)^2}.
\end{align}
\\(c) Magnetohydrodynamic (MHD) turbulence
\begin{align}
\label{eq23}
\Omega _{{\text{gw}}*}^{{\text{MHD}}}(\nu ) = \left( {\frac{{{H_*}}}{\beta }} \right){\chi _{{\text{MHD}}}}{S_{{\text{MHD}}}},
{\chi _{{\text{MHD}}}} = {\left( {\frac{{{\kappa _{{\text{MHD}}}}}}{{1 + \epsilon}}} \right)^{\frac{3}{2}}}.
\end{align}
In Eq.~(\ref{eq21})-Eq.~(\ref{eq23}), the $\beta$  is the inverse time duration of the phase transition,  ${H_*}$ is the Hubble parameter at the time of production of GWs,  $\mu$ is the velocity of wall, $\epsilon$  denotes the ratio of the vacuum energy density released in the phase transition to that of the radiation, ${\kappa _{{\text{BC}}}}$,  ${\kappa _{{\text{SW}}}}$ and ${\kappa _{{\text{MHD}}}}$  are represent the fraction of the latent heat of the phase transition deposited on the BC, SW, and MHD turbulence, respectively. The spectral functions of the SGWB which are characterized from numerical f\/its as
\begin{subequations}
\begin{align}
\label{eq23+}
{S_\text{BC}} = \frac{{3.8{{\left( {{\nu  \mathord{\left/  {\vphantom {\nu  {{\nu _\text{BC}}}}} \right. \kern-\nulldelimiterspace} {{\nu _\text{BC}}}}} \right)}^{2.8}}}}{{1 + 2.8{{\left( {{\nu  \mathord{\left/ {\vphantom {\nu  {{\nu _\text{BC}}}}} \right. \kern-\nulldelimiterspace} {{\nu _\text{BC}}}}} \right)}^{3.8}}}},
\end{align}
\begin{align}
\label{eq23++}
{S_{\text{SW}}} = {\left( {\frac{\nu }{{{\nu _\text{SW}}}}} \right)^3}{\left[ {\frac{7}{{4 + 3{{\left( {{\nu  \mathord{\left/ {\vphantom {\nu  {{\nu _\text{SW}}}}} \right.
 \kern-\nulldelimiterspace} {{\nu _\text{SW}}}}} \right)}^2}}}} \right]^{3.5}},
 \end{align}
 \begin{align}
\label{eq24}
{S_{{\text{MHD}}}} = \frac{{\mu {{\left( {\frac{\nu }{{{\nu _{{\text{MHD}}}}}}} \right)}^3}}}{{{{\left( {1 + \frac{\nu }{{{\nu _{{\text{MHD}}}}}}} \right)}^{\frac{{11}}{3}}}\left\{ {1 + \frac{{8\pi \nu }}{{{H_*}}}{{\left[ {\frac{{{T_0}}}{{{T_*}}}\frac{{{g_s}{{\left( {{T_0}} \right)}^{\frac{{2 + \delta }}{6}}}}}{{{g_s}{{\left( {{T_*}} \right)}^{\frac{{2 + \delta }}{6}}}}}{{\left( {\frac{{{T_0}}}{{{T_*}}}} \right)}^{\frac{\delta }{2}}}} \right]}^{ - 1}}} \right\}}},
\end{align}
\label{eq23-a}
\end{subequations}
with the peak frequencies as
\begin{subequations}
\begin{align}
\label{eq26}
{\nu _\text{BC}} & = \frac{{0.62\beta }}{{1.8 - 0.1\mu  + {\mu ^2}}}\left( {\frac{{{a_*}}}{{{a_0}}}} \right)
\nonumber \\
& = \frac{{0.62\beta }}{{1.8 - 0.1\mu  + {\mu ^2}}}\frac{{{T_0}}}{{{T_*}}}\frac{{{g_s}{{\left( {{T_0}} \right)}^{\frac{{2 + \delta }}{6}}}}}{{{g_s}{{\left( {{T_*}} \right)}^{\frac{{2 + \delta }}{6}}}}}{\left( {\frac{{{T_0}}}{{{T_*}}}} \right)^{\frac{\delta }{2}}},
\end{align}
\begin{align}
\label{eq27}
{\nu _\text{SW}} = \frac{{38\beta }}{{31\mu }}\left( {\frac{{{a_*}}}{{{a_0}}}} \right) = \frac{{38\beta }}{{31\mu }}\frac{{{T_0}}}{{{T_*}}}\frac{{{g_s}{{\left( {{T_0}} \right)}^{\frac{{2 + \delta }}{6}}}}}{{{g_s}{{\left( {{T_*}} \right)}^{\frac{{2 + \delta }}{6}}}}}{\left( {\frac{{{T_0}}}{{{T_*}}}} \right)^{\frac{\delta }{2}}},
\end{align}
\begin{align}
\label{eq28}
{\nu _{{\text{MHD}}}} & = \frac{{7\beta }}{{4\mu }}\left( {\frac{{{a_*}}}{{{a_0}}}} \right)
\nonumber \\
& = \frac{{7\beta }}{{4\mu }}\left[ {\frac{{{T_0}}}{{{T_*}}}\frac{{{g_s}{{\left( {{T_0}} \right)}^{\frac{{2 + \delta }}{6}}}}}{{{g_s}{{\left( {{T_*}} \right)}^{\frac{{2 + \delta }}{6}}}}}{{\left( {\frac{{{T_0}}}{{{T_*}}}} \right)}^{\frac{\delta }{2}}}} \right].
\end{align}
\label{eq24-a}
\end{subequations}
It is should be noted that the parameters  $\epsilon$ and $\kappa$  in $\chi$  play essential roles in the def\/inition of the peak position and amplitude of the SGWB signal, nevertheless, since they are model-dependent, there is still no solid method to find a specific expression for  $\kappa$. According to the viewpoints in~\cite{cha19,cha39}, we set  ${\kern 1pt} {\kern 1pt} {\chi _{\text{BC}}} = {\kern 1pt} {\kern 1pt} {\chi _{\text{SW}}} = {\kern 1pt} {\kern 1pt} {\chi _{\text{MHD}}} = 0.05$, $\beta  = n{H_*}$,  $n=5$,  $\mu=0.7$, and the Hubble parameter at phase transition epoch can be expressed as
\begin{align}
\label{eq28}
{H_*} = \sqrt {\frac{{8\pi }}{{3m_p^2}}\rho \left( {{T_*}} \right)} ,
\end{align}
where $m_p$  is the Planck mass, $\rho \left( {{T_*}} \right) = T_*^5\left[ {{{{\text{d}}F\left( {{T_*}} \right)} \mathord{\left/ {\vphantom {{{\text{d}}F\left( {{T_*}} \right)} {{\text{d}}{T_*}}}} \right.  \kern-\nulldelimiterspace} {{\text{d}}{T_*}}}} \right] + 3T_*^4F\left( {{T_*}} \right)$  is the energy density at transition temperature  ${T_*}$. It is well known that the QCD phase transitions occurs at temperatures ranges from about a few hundred MeV, whose exact value depends on the type of QCD matters. For the sake of simplicity, here we assume that GW is generated immediately after the phase transition occurrence, hence, the temperature ${T_*}$  is estimated to be the critical phase change temperature  ${T_c}$, that is,  ${T_*} = {T_c} = 0.145$  GeV. As a result, according to the expressions of today's peak frequency of the SGWB generated BC, SW, MHD turbulence, the total peak frequency of SGWB in the Barrow entropy background is given by
\begin{align}
\label{eq29}
  {\nu _{{\text{total}}}} & = \left( {\frac{{0.62\beta }}{{1.8 - 0.1\mu  + {\mu ^2}}} + \frac{{38\beta }}{{31\mu }} + \frac{{7\beta }}{{4\mu }}} \right)\left( {\frac{{{a_*}}}{{{a_0}}}} \right)
\nonumber \\
&   = \left( {\frac{{0.62\beta }}{{1.8 - 0.1\mu  + {\mu ^2}}} + \frac{{38\beta }}{{31\mu }} + \frac{{7\beta }}{{4\mu }}} \right){\left[ {\frac{{{g_s}\left( {{T_0}} \right)}}{{{g_s}\left( {{T_*}} \right)}}} \right]^{\frac{{2 + \delta }}{6}}}
\nonumber \\
& \times {\left( {\frac{{{T_0}}}{{{T_*}}}} \right)^{1 + \frac{\delta }{2}}}.
\end{align}

\begin{figure*}[htbp]
\centering
\subfigure[]{
\includegraphics[scale=0.31]{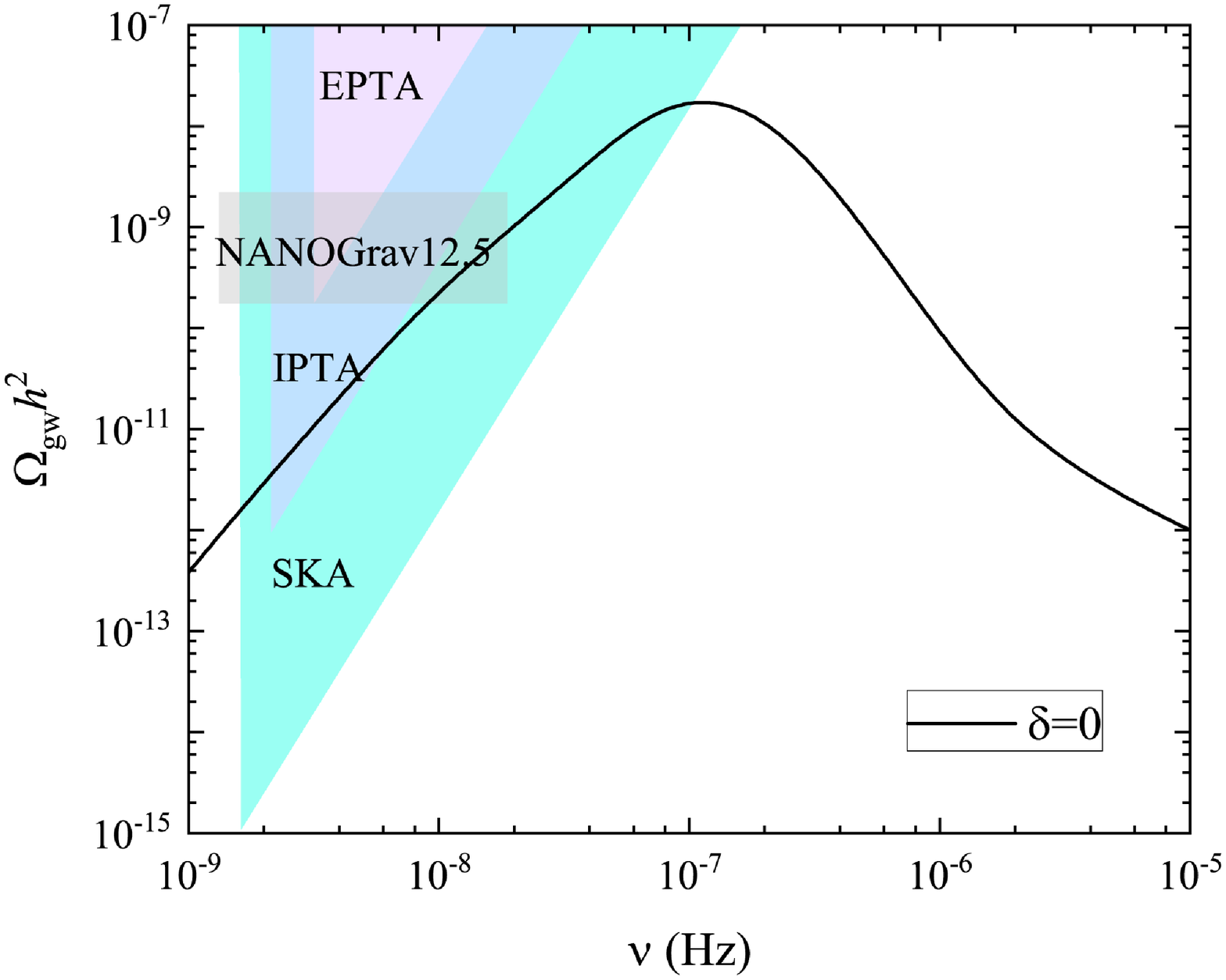}
\label{fig6-a}
}
\quad
\subfigure[]{
\includegraphics[scale=0.31]{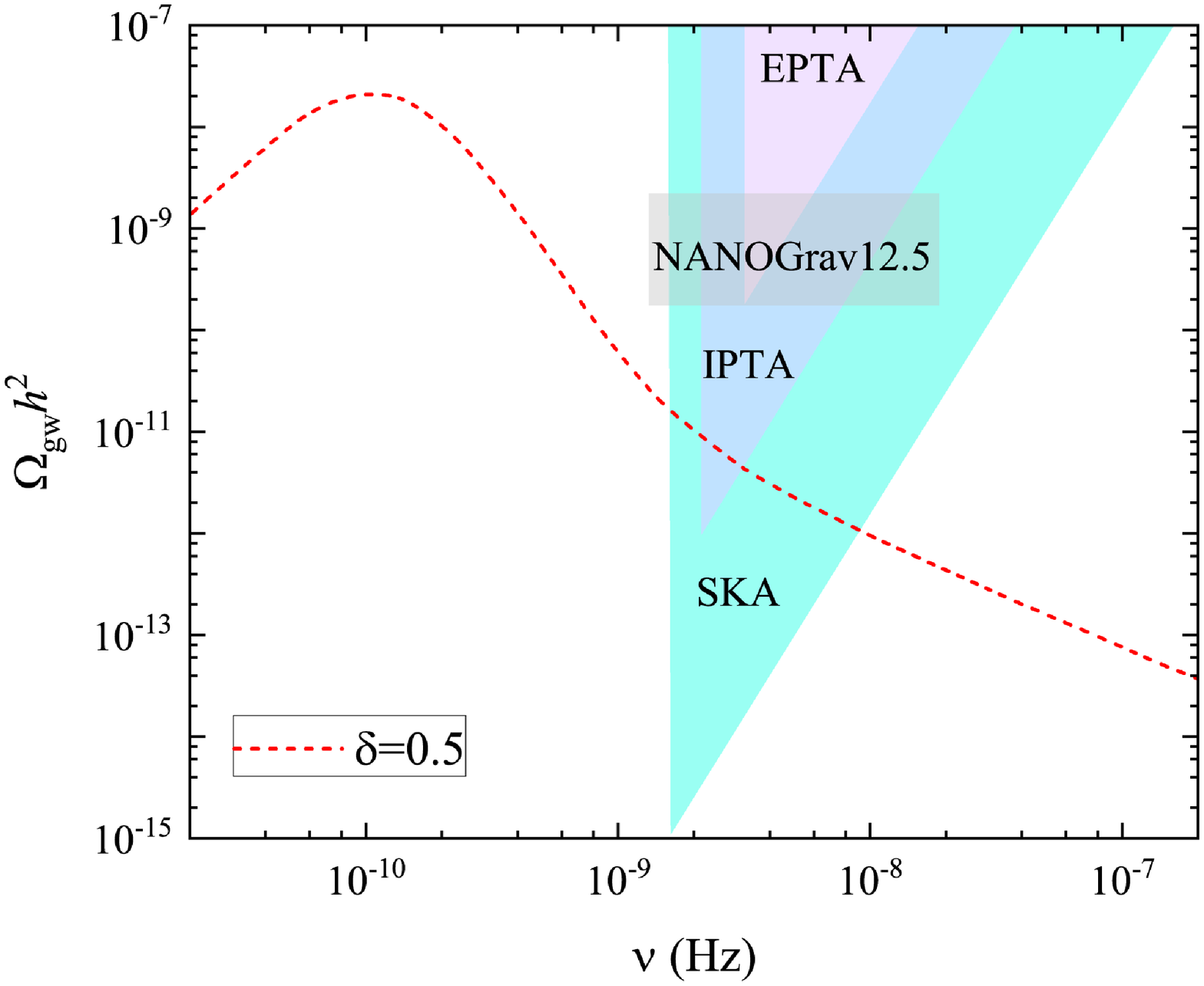}
\label{fig6-b}
}
\quad
\subfigure[]{
\includegraphics[scale=0.31]{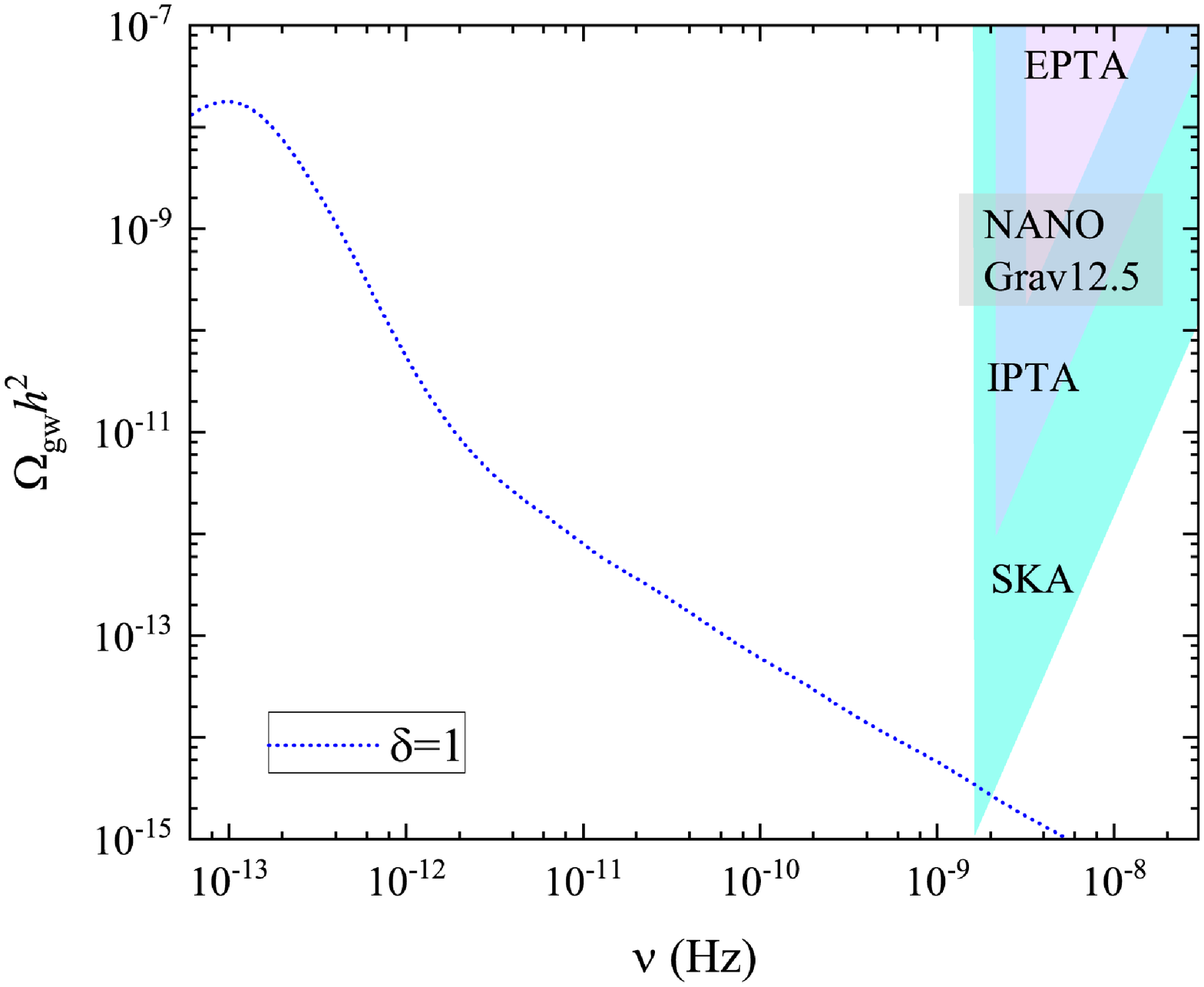}
\label{fig6-c}
}
\quad
\subfigure[]{
\includegraphics[scale=0.31]{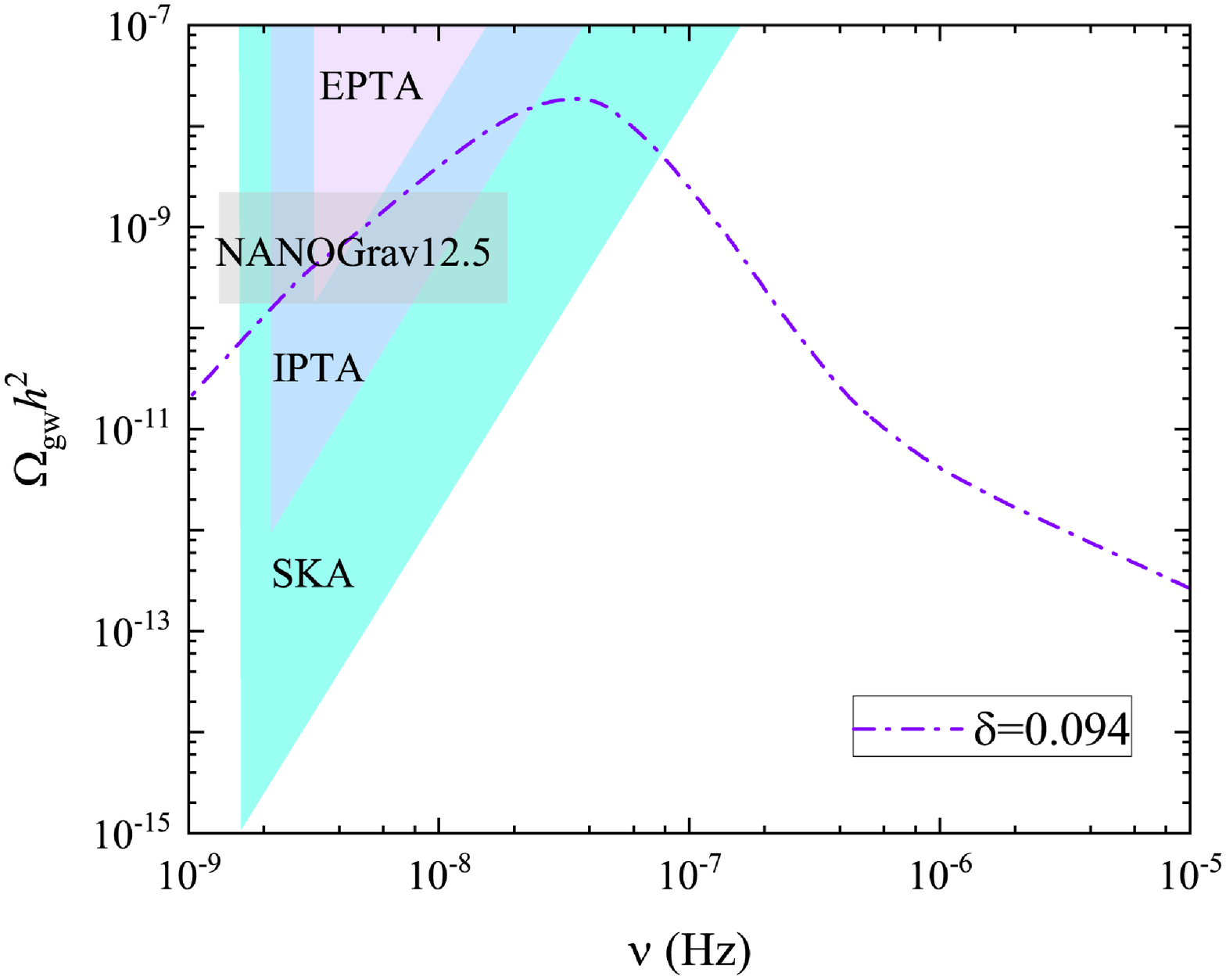}
\label{fig6-d}
}
\caption{The SGWB spectrum from QCD phase transition as a function of frequency for dif\/ferent Barrow exponent  $\delta$.}
\label{fig6}
\end{figure*}

Finally, based on Eq.~(\ref{eq20})-Eq.~(\ref{eq28}), the total energy density of the SGWB can be expressed as follows
\begin{align}
\label{eq30}
  {\Omega _{{\text{gw}}}}{h^2} & = \left[ {\Omega _{{\text{gw}}}^{BC}\left( \nu  \right) + \Omega _{{\text{gw}}}^{SW}\left( \nu  \right) + \Omega _{{\text{gw}}}^{{\text{MHD}}}\left( \nu  \right)} \right]{h^2}
\nonumber \\
 &= \left[ {\Omega _{\text{gw}*}^{BC}\left( \nu  \right) + \Omega _{\text{gw}*}^{SW}\left( \nu  \right) + \Omega _{{\text{gw}}*}^{{\text{MHD}}}\left( \nu  \right)} \right] {\Omega _{r0}}{h^2}
\nonumber \\
& \times {\left( {\frac{{{T_r}}}{{{T_*}}}} \right)^{2\left( {2 + \delta } \right)}}{\left[ {\frac{{{g_s}\left( {{T_r}} \right)}}{{{g_s}\left( {{T_*}} \right)}}} \right]^{\frac{2}{3}\left( {2 + \delta } \right)}} {\left[ {\frac{{{g_s}{{\left( {{T_*}} \right)}^{1 + \omega \left( {{T_*}} \right)}}}}{{{g_s}{{\left( {{T_r}} \right)}^{1 + \omega \left( {{T_r}} \right)}}}}} \right]^{\frac{{2 + \delta }}{2}}}
\nonumber \\
& \times\exp \left[ {\int_{{T_r}}^{{T_*}} {\frac{3}{T}\left( {1 + {\omega _{{\text{ef\/f}}}}} \right)\left( {1 + \frac{\delta }{2}} \right){\text{d}}T} } \right],
\end{align}
where $h = {{{H_0}} \mathord{\left/ {\vphantom {{{H_0}} {100{\text{k}}{{\text{m}}^{ - 1}}}}} \right. \kern-\nulldelimiterspace} {100{\text{k}}{{\text{m}}^{ - 1}}}}{\text{s Mpc}}$. Substituting the parameters mentioned above into Eq.~(\ref{eq30}), we show the SGWB spectrum from QCD phase transitions due to BC, SW and  MHD turbulence for dif\/ferent parameter $\delta$ in Fig.~\ref{fig6}.

From the subfigures in Fig.~\ref{fig6}, one can see that the effect of Barrow entropy suppresses the frequency of the SGWB signal $\nu$  but has no signif\/icant inf\/luence on its energy density $ {\Omega _{{\text{gw}}}}{h^2} $.  With these results, one can explore the possibility of the detection of  SGWB generated from the f\/irst-order cosmological QCD phase in the framework of Barrow entropy. In Fig.~\ref{fig6-a}, the black solid curve represents the original SGWB spectrum without Barrow entropy, whose peak frequency of the SGWB is around $100$ nHz. In addition, the curve lies within the sensitive region of SKA, IPTA, and NANOGrav 12.5-yr observation, which implies the SGWB signal can be detected by SKA, IPTA, and NANOGrav. However, when considering the ef\/fect of Barrow entropy, it is clear that the SGWB spectrum moves to the lower frequency region. From Fig.~\ref{fig6-b}, one can see that the peak frequency of the SGWB with $\delta=0.5$  drops to around  $10^{-1}$ nHz, which results in its signal being detected only by SKA and EPTA.  In the limiting case $\delta=1$, the Fig.~\ref{fig6-c} shows  the peak frequency of the SGWB reduces to around $10^{-4}$ nHz, and the probability of detection of the SGWB signal becomes very low since only the tail of the signal reaches the projection sensitivity of the SKA. As a result, the signal of SGWB with  $\delta=1$ is the most challenging to detect, which require more sensitive detectors for discovery.

In light of the importance role of Barrow entropy for the study of the properties of  black holes and the early  universe, an increasing number of investigations in recent years have been carried out  to constrain the range of Barrow exponent  $\delta$. According to the observational date of Supernovae (SNIa) and Hubble parameter, Anagnostopoulos, Basilakos and Saridakis \cite{cha42} constrained on the scenario of Barrow holographic dark energy found outcomes of  $\delta  = 0.094_{ - 0.101}^{ + 0.094}$. By using  the Big Bang Nucleosynthesis (BBN) data, the upper bound of the exponent of Barrow entropy is constrained to be  $\delta  \leq 1.4 \times {10^{ - 4}}$ \cite{cha43}.  Furthermore, the data from M87* central black hole shadow and the S2 star observations have been used to extract constrained on Barrow exponent and the result showed that  $\delta  \leq 0.0828$ at 1$\sigma$  \cite{cha44}. More recently, based on the modified Friedman equations due to Barrow entropy and the gravitational baryogenesis, the authors in~\cite{cha45} pointed out that the range of Barrow exponent should be $0.005 \leq \delta  \leq 0.008$.  According to the
above results, e.g.,  $\delta  = 0.094$, we can analyze the observability of SGWB in the framework of Barrow entropy. From Fig.~\ref{fig6-d}, one can see that the peak frequency of SGWB has shifted left to  $\nu  = 3 \times {10^{ -8 }}$ Hz and the SGWB signal (purple dashed-dotted curve) exceed the sensitivity curve of  SKA, IPTA, EPTA and NANOGrav 12.5-yr observation, which indicates that the SGWB signal in the formwork of Barrow entropy  can be detected by detectors such as the SKA, IPTA, EPTA, and NANOGrav.

Now, it is necessary to discuss our results. Firstly, it is clear that the presence of the ef\/fect of Barrow entropy has an important impact on the properties of the SGWB. Then, the SGWB signals with dif\/ferent Barrow  exponent could be discovered by dif\/ferent detectors. In particular, when applying the most desirable result $\delta  = 0.094$, it is found that the SGWB signal can be detected by SKA, IPTA, EPTA, and NANOGrav. Notably, most constraint works (BBN, the data from M87* central black hole shadow and S2 star observations, the modif\/ied Friedman equations and gravitational baryogenesis) on the Barrow exponent showed the deviations of the $\delta$ from the classical result 0 is very small, which is consistent with the expectation of $\delta$ in a realistic case. Therefore, the SKA is most likely to detect the SGWB signal since its detection sensitivity is the highest  in the current stage. Nevertheless, the results given by only one detector are not convincing enough. To further conf\/irm the existence of the Barrow entropy ef\/fect, as well as the QG and the fractal structure of the universe it entail, more precise experiments are needed. Fortunately, with the increased attention to the problem of GWs, more detectors will be operating in the near future, which will help to solve these problems.

\section{Conclusion}
\label{sec5}
In this paper, in a framework where Barrow entropy  are present, we investigated the SGWB generated by the f\/irst-order cosmological QCD phase transition of the  early univers. This work started with the assumption the universe has expanded adiabatically, in this case the effect of Barrow entropy can be found to be signif\/icantly inf\/luence the time variation of universe temperature  ${{{\text{d}}T} \mathord{\left/ {\vphantom {{{\text{d}}T} {{\text{d}}t}}} \right. \kern-\nulldelimiterspace} {{\text{d}}t}}$, the relation between the scale factor and redshift  ${{{{a_*}} \mathord{\left/ {\vphantom {{{a_*}} {{a_0}}}} \right.  \kern-\nulldelimiterspace} {{a_0}}}}$, the redshift in the SGWB frequency peak relative to the corresponding value at current epoch  $ {{{{\nu _{0{\text{peak}}}}} \mathord{\left/ {\vphantom {{{\nu _{0{\text{peak}}}}} {{\nu _*}}}} \right. \kern-\nulldelimiterspace} {{\nu _*}}}}$, as well as the density parameter of the gravitational waves at current epoch ${\Omega _{{\text{gw}}}}$. Then, based on recent lattice calculations, we analyzed the impact of the trace anomaly on the equation of state during the QCD phase transition and argued for itsnon-negligible contribution to the generation of the SGWB signal. Finally, by taking account of Bubble wall collisions, sound waves and magnetohydrodynamic turbulence as the sources of SGWB,  an analysis of the inf\/luence of Barrow entropy on the total energy density and the peak signal of SGWB signal is carried out. Our results showed that the effect of Barrow entropy can suppress ${{{{\nu _{0{\text{peak}}}}} \mathord{\left/ {\vphantom {{{\nu _{0{\text{peak}}}}} {{\nu _*}}}} \right. \kern-\nulldelimiterspace} {{\nu _*}}}}$, the ratio of SGWB spectrum measured today to the time of phase transition ${{{\Omega _{{\text{gw}}}}} \mathord{\left/  {\vphantom {{{\Omega _{{\text{gw}}}}} {{\Omega _{{\text{gw}}*}}}}} \right. \kern-\nulldelimiterspace} {{\Omega _{{\text{gw}}*}}}}$, and the total peak signal of SGWB today ${\nu _{{\text{total}}}}$, but increase the evaluation of Hubble parameter from phase transition epoch to the current time ${{{H_*}} \mathord{\left/ {\vphantom {{{H_*}} {{H_0}}}} \right. \kern-\nulldelimiterspace} {{H_0}}}$. Importantly, we f\/ind that ef\/fect of Barrow entropy reduce the frequency of the SGWB signal, so that dif\/ferent detectors (SKA, IPTA, NANOGrav, and EPTA) can detect gravitational wave signals with dif\/ferent  $\delta$.  Furthermore, according to the previous constrained works on the Barrow exponent $\delta$, we suppose that SGWB with $\delta=0.094$ can be detected at least by SKA, IPTA, NANOGrav and EPTA. This in turn helps people study the properties of QG and the fractal structure of the universe.

Aside from the above results, it is found that the ef\/fect of Barrow entropy could be equivalently parametrized by dilution effect $D$ in \cite{chc5}, which can be interpreted as a form of dark radiation. Besides, since the e\/ffect of  Tsallis entropy could lead to cosmological implications \cite{cha46,cha47}, it is also  interesting to investigate the SGWB in the framework of Tsallis entropy. We hope to report in future studies.

\vspace*{0.2ex}
{\bf Acknowledgements}

We thank Mohamed Moussa for his enthusiastic help.

%\vspace*{3.0ex}
%{\bf Acknowledgements}
%\vspace*{1.0ex}

\end{document}